**Paper title**

Diversity in immunogenomics: the value and the challenge

**Authors and affiliations**

Kerui Peng
Department of Clinical Pharmacy, School of Pharmacy, University of Southern California, CA, USA
keruipen@usc.edu

Yana Safonova
Computer Science and Engineering Department, University of California San Diego, San Diego, USA, CA; Department of Biochemistry and Molecular Genetics, University of Louisville School of Medicine, Louisville, KY, USA
isafonova@eng.ucsd.edu

Mikhail Shugay
Shemyakin-Ovchinnikov Institute of Bioorganic Chemistry, Russian Academy of Sciences, Moscow, Russia
Pirogov Russian National Research Medical University, Moscow, Russia
mikhail.shugay@gmail.com

Alice Popejoy
Department of Biomedical Data Science (DBDS), Stanford University, Stanford, CA, USA
apopejoy@stanford.edu

Oscar L. Rodriguez
Department of Biochemistry and Molecular Genetics, University of Louisville School of Medicine, Louisville, KY, USA
oscar.rodriguez.1@louisville.edu

Felix Breden
Department of Biological Sciences, Simon Fraser University, Burnaby bc v5a 1s6 Canada
breden@sfu.ca

Petter Brodin
Science for Life Laboratory, Dept. of Women's and Children Health, Karolinska Institutet, Sweden
Pediatric Rheumatology, Karolinska University Hospital, Sweden

petter.brodin@ki.se

Amanda M. Burkhardt
Department of Clinical Pharmacy, School of Pharmacy, University of Southern California, CA, USA
aburkhar@usc.edu

Carlos Bustamante
Department of Biomedical Data Science (DBDS), Stanford University, Stanford, CA, USA
cdbustam@stanford.edu

Van-Mai Cao-Lormeau
Laboratory of research on Infectious Vector-borne diseases, Institut Louis Malardé, French Polynesia
mlormeau@ilm.pf

Martin M. Corcoran
Department of Microbiology, Tumor and Cell Biology, Karolinska Institutet, Sweden
martin.corcoran@ki.se

Darragh Duffy
Translational Immunology Lab, Institut Pasteur, 25 Rue du Dr Roux, Paris, France
darragh.duffy@pasteur.fr

Macarena Fuentes Guajardo
Health Sciences Faculty, University of Tarapacá, Arica, Chile
macarena.fuentes.14@alumni.ucl.ac.uk

Ricardo Fujita
Centro de Genética y Biología Molecular, Universidad de San Martín de Porres, Alameda del Corregidor 1531, La Molina, Lima 12, Perú
rfujitaa@usmp.pe

Victor Greiff
Department of Immunology, University of Oslo, Oslo, Norway
victor.greiff@medisin.uio.no

Vanessa D. Jönsson
Departments of Computational and Quantitative Medicine and Hematology, Beckman Research Institute, City of Hope, Duarte, CA, USA
vjonsson@coh.org

Xiao Liu
School of Life Sciences, Tsinghua University, Beijing, China
airson_liu@163.com

Lluis Quintana-Murci
Human Evolutionary Genetics Unit, CNRS UMR2000, Institut Pasteur, 25-28 rue du Dr Roux, Paris, France
Chair of Human Genomics and Evolution, Collège de France, 11 Place Marcelin Berthelot, Paris, France
lluis.quintana-murci@pasteur.fr

Maura Rossetti
UCLA Immunogenetics Center, Department of Pathology and Laboratory Medicine, David Geffen School of Medicine, University of California Los Angeles, Los Angeles, CA, USA
maurarossetti@gmail.com

Jianming Xie
Department of Pharmacology & Pharmaceutical Sciences, School of Pharmacy, University of Southern California, CA, USA
jianminx@usc.edu

Gur Yaari
Faculty of Engineering, Bar Ilan Institute of Nanotechnologies and Advanced Materials, Bar Ilan University, Ramat Gan, Israel
gur.yaari@biu.ac.il

Wei Zhang
Department of Computer Science, City University of Hong Kong, Hong Kong, China
wzhang287-c@my.cityu.edu.hk

Malak S. Abedalthagafi
Genomics Research Department, Saudi Human Genome Project, King Fahad Medical City and King Abdulaziz City for Science and Technology, Riyadh, Saudi Arabia
malthagafi@kacst.edu.sa

Khalid O. Adekoya
Department of Cell Biology and Genetics, University of Lagos, Lagos, Nigeria
kadekoya@unilag.edu.ng

Rahaman A. Ahmed
Department of Cell Biology and Genetics, University of Lagos, Lagos, Nigeria
ahmedrahman158@gmail.com

Wei-Chiao Chang
Department of Clinical Pharmacy, School of Pharmacy, Taipei Medical University, Taiwan
Division of Nephrology, Department of Internal Medicine, Taipei Medical University-Shuang Ho Hospital, New Taipei City, Taiwan
wcc@tmu.edu.tw

Clive Gray
Division of Immunology, Institute of Infectious Disease and Molecular Medicine and Department of Pathology, University of Cape Town
Laboratory of Tissue Immunology, National Health Laboratory Services and Groote Schuur Hospital
clive.gray@uct.ac.za

Yusuke Nakamura
Cancer Precision Medicine Center, Japanese Foundation for Cancer Research
yusuke.nakamura@jfcr.or.jp

William D. Lees
Institute of Structural and Molecular Biology, Birkbeck College, London, UK
william@lees.org.uk

Purvesh Khatri
Institute for Immunity, Transplantation and Infection, School of Medicine, Stanford University, CA, USA
Center for Biomedical Research, Department of Medicine, School of Medicine, Stanford University, CA, USA
pkhatri@stanford.edu

Houda Alachkar†
Department of Clinical Pharmacy, School of Pharmacy, University of Southern California, CA, USA
alachkar@usc.edu

Cathrine Scheepers†

Centre for HIV and STIs, National Institute for Communicable Diseases of the National Health Laboratory Service and the South African Medical Research Council Antibody Immunity Research Unit, School of Pathology, University of the Witwatersrand, Johannesburg, South Africa
cathrinem@nicd.ac.za

Corey T. Watson†
Department of Biochemistry and Molecular Genetics, University of Louisville School of Medicine, Louisville, KY, USA
corey.watson@louisville.edu

Gunilla B. Karlsson Hedestam†
Department of Microbiology, Tumor and Cell Biology, Karolinska Institutet, 171 77 Stockholm, Sweden
gunilla.karlsson.hedestam@ki.se

Serghei Mangul†
Department of Clinical Pharmacy, School of Pharmacy, University of Southern California, CA, USA
serghei.mangul@gmail.com
ORCID: https://orcid.org/0000-0003-4770-3443
Twitter: smangul1

*Correspondence: serghei.mangul@gmail.com (S. Mangul)
†These authors contributed equally to the work.

**Abstract**

With the advent of high-throughput sequencing technologies, the fields of immunogenomics and adaptive immune receptor repertoire research are facing both opportunities and challenges. Adaptive immune receptor repertoire sequencing (AIRR-seq) has become an increasingly important tool to characterize T and B cell responses in settings of interest. However, the majority of AIRR-seq studies conducted so far were performed in individuals of European ancestry, restricting the ability to identify variation in human adaptive immune responses across populations and limiting their applications. As AIRR-seq studies depend on the ability to assign VDJ sequence reads to the correct germline gene segments, efforts to characterize the genomic loci that encode adaptive immune receptor genes in different populations are urgently needed. The availability of comprehensive germline gene databases and further applications of AIRR-seq studies to individuals of non-European ancestry will substantially enhance our understanding of human adaptive immune responses, promote the development of effective diagnostics and treatments, and eventually advance precision medicine.

**Main**

*Diversity in genomics studies and AIRR-seq*

So far, genomic studies have mainly used samples from individuals of European ancestry, at the expense of learning from the largest and most genetically diverse populations. For example, 78% of individuals included in genome-wide association studies (GWAS) reported in the GWAS Catalog (https://www.ebi.ac.uk/gwas/home) through January 2019 are of European descent[1], while Asian populations account for 59.5% of the world population based on the Population Reference Bureau's World Population Data Sheet (https://www.prb.org/datasheets/). Though partially due to inadequate sampling of non-European populations, even when diverse datasets are available, researchers tend to exclude data from minority groups when conducting statistical analyses[2].

In recent years, there has been an increased awareness of the limited generalizability of findings across populations, motivating the inclusion of diverse, multiethnic populations in large-scale genomic studies[3,4]. Whole-genome sequencing in individuals of African descent[5–8] and whole-exome sequencing in a southern African population[9] have improved understanding of genetic variation in underrepresented populations. Additional efforts have been made to establish reference genome datasets for research in diverse populations; including the GenomeAsia 100K Project[10], the Human Heredity and Health in Africa (H3Africa) initiative[11,12], the Taiwan Biobank[13], Population Architecture Using Genomics and Epidemiology (PAGE) Consortium[14], Trans-Omics for Precision Medicine (TOPMed) program, Clinical Sequencing Evidence-Generating Research (CSER) consortium[15], Human Genome Reference Program (HGRP), and All of Us Research Program[16].

The inclusion of diverse populations in genomic studies has demonstrated benefits for the discovery and interpretation of gene-trait associations. Similarly, greater diversity in immunogenomics research and AIRR-seq will enable the discovery of novel genetic traits associated with immune system phenotypes that are common or different across populations. Broader inclusion of diverse populations may also enable researchers to address genetic heterogeneity in the context of translational research, possibly revealing clinically relevant genomic signatures that are more prevalent in some populations than others.

Central to immunity are the repertoires of T cell receptors (TCRs), immunoglobulins (IGs), human leukocyte antigens (HLAs) and killer cell immunoglobulin-like receptors (KIRs). Thus, analyses of the loci that encode these molecules are critical to immunogenomics studies. There have been extensive efforts to address the genetic diversity of the HLA and KIR systems[17,18]. However, the current understanding of diversity in the TCR and IG loci is more limited. We therefore focus the discussion on TCR and IG germline gene diversity here.

T cells and B cells recognize antigens through their TCRs and IGs, which are formed through the process of V(D)J recombination. Capturing the vast diversity of recombined, expressed TCR and IG repertoires was not possible until the development of high-throughput sequencing techniques in the late 2000s. Freeman and colleagues were the first to employ 5′ rapid amplification of cDNA end (5′RACE) PCR to amplify TCR cDNA and to characterize TCR repertoires[19]. Additionally, Weinstein and colleagues sequenced the first antibody repertoire in zebrafish in 2009[20], creating the foundation of AIRR-seq technologies. In 2010, the initial application of AIRR-seq technology in human IGs was made by Boyd and colleagues[21]. Since then, studies including AIRR-seq have seen exponential growth, and findings from these studies have shaped our current understanding of human immune repertoires in different settings[22–25]. AIRR-seq analysis and other immunogenomics studies offer new opportunities to deepen our understanding of the immune system in the context of a variety of human diseases, including infectious diseases[26–30], cancer[31,32], autoimmune conditions[33–36], and neurodegenerative disease[37]. Furthermore, AIRR-seq data provides information on expression profiles, germline V, D, and J gene usage, complementarity determining region (CDR) diversity, and somatic hypermutation (SHM) levels.

*Germline gene diversity and databases*

A critical step in AIRR-seq studies is germline gene assignment, which requires reliable and comprehensive databases of germline VDJ alleles. So far, such databases are lacking since the genetic regions encoding these genes have been exceptionally challenging to characterize at the genomic level. Not only do these loci contain a mixture of functional genes and pseudogenes with high similarity, but they are also characterized by a considerable level of structural variation with deletions and duplications occurring at high frequency in different populations. Given the complexity of the TCR and IG genomic loci, and deficits in existing germline databases, the determination of germline immune receptor gene usage from bulk RNA-seq or whole genome sequencing is often inaccurate. Efforts to improve germline databases are therefore critical, not the least for improved coverage of diversity in immune repertoire analysis. Computational methods to infer germline TCR and IG genes from AIRR-seq data are expected to accelerate these efforts[38–43] (Supplemental Table 1). Additionally, comparisons are needed between results obtained from methods for inferring germline gene variants from AIRR-seq repertoires[44] and from direct sequencing of genomic DNA[21,45–47], such as the sequencing and assembly of large-insert clones (e.g., bacterial artificial chromosome (BAC) and fosmid clones)[48], and more recently whole-genome sequencing (WGS)[49] and targeted long read sequencing[50].

The most widely used reference database for immunogenomics data, the international ImMunoGeneTics information system (IMGT)[51], has historically been a valuable resource. However, it lacks a comprehensive set of human TCR and IG alleles representing diverse populations worldwide. Additional uncertainty stems from the fact that descriptions of sample populations in databases are often self-identified based on geography or ethnicity, rather than genetic ancestry. As a result, we have a limited understanding of population-level differences in germline variation. However, progress is being made to address these issues. The AIRR Community (AIRR-C; www.airr-community.org) is an international community of bioinformaticians and immunogeneticists that formed to develop standards and protocols to promote sharing and common analysis approaches for AIRR-seq data, including the AIRR Data Commons [52]. As a means to enrich available germline gene sets, the AIRR-C established the Inferred Allele Review Committee (IARC; https://www.antibodysociety.org/the-airr-community/airr-subcomittees/inferred-allele-review-committee-iarc) to review and curate previously unrecorded IG/TCR germline genes inferred from AIRR-seq data. Its work is underpinned by the Open Germline Receptor Database (OGRDB), which provides submission and review workflows. IARC-affirmed sequences are published in OGRDB, together with supporting evidence. VDJbase was also recently launched as a public database that allows users to access population-level IG/TCR germline data, including reports and summary statistics on germline genes, alleles, single nucleotide and structural variants, and haplotypes of interest derived from AIRR-seq and genomic sequencing data[53]. It currently contains AIRR-seq data from 560 human donors, representing 654 immunoglobulin heavy and light chain gene alleles. The integration of TCR datasets is currently in progress. Together these initiatives will help pave the way for the development of approaches that extend germline curation efforts to include more data types and ultimately ensure that population-level metadata can be more effectively captured and leveraged.

Population genetic differences have been observed in genomics studies and immunogenomics is no exception[45,54–57]. For example, evidence for extensive diversity in germline TCR and IG genes have been reported in the human population[46–48,55,56,58]. However, it is notable that most AIRR-seq studies have been conducted in individuals of European descent, leaving other populations underrepresented[59,60]. The limited inclusion of samples from diverse populations hinders the advancement of many areas of medicine as it leaves uncertainty with respect to the genetic basis of treatment outcomes and health disparities[60]. Moreover, it limits our understanding of how pathogens have exerted selective pressures on immune-related genes in populations living in different environments, and thus on infectious disease manifestation[61]. We argue that these shortcomings must be addressed through efforts that seek to include more diverse populations in immunogenomics research.

As an interdisciplinary group, with expertise in biomedical and translational research, population biology, computational biology, and immunogenomics, we wish to raise awareness about the value of including diverse populations in AIRR-seq and immunogenomics research. In the areas of genetic disease research and cancer genomics, enhanced genetic diversity has led to demonstrable insights[62,63]. However, the field of immunogenomics has yet to benefit from a similar growth in diversity. At the current stage of the global COVID-19 pandemic, numerous vaccine trials and programs are underway worldwide, offering opportunities to investigate the role of genetic factors on vaccine mediated immune responses. Such studies will require careful study designs to effectively address potential confounding factors such as environmental and socio-economic differences[64,65]. Incomplete representation of diverse populations limits our capacity to address the impact of genetics on clinical phenotypes, and ideally this should be investigated alongside environmental risk factors for disease. Previous studies have shown associations between genetic variants and clinical presentation of disease[66,67]. Specific IG germline genes, and in some cases alleles, have been found to be preferentially used in response to pathogens such as influenza[26], HIV-1[68], Zika[69], and SARS CoV-2[70]. Vaccine effectiveness and infection outcomes are likely shaped by genetic variability, including specific effects driven by immune genes[60,71].

We propose that the community should make efforts to include underrepresented populations in AIRR and immunogenomics studies. Already, those that have conducted AIRR-seq in populations of non-European descent have uncovered evidence for extensive germline diversity. For example, in a study of South African HIV patients, Scheepers and colleagues discovered a large number of immunoglobulin heavy chain variable (IGHV) alleles that were not represented in IMGT[46], information of relevance to HIV vaccine design aimed at germline targeting immunogens[72]. In a study in the Papua New Guinea population, one undocumented IGHV gene and 16 IGHV allelic variants were identified from AIRR-seq data[47]. These discoveries of undocumented alleles indicate the need to generate population-based AIRR-seq datasets and to identify and validate the presence of undocumented alleles such that they can be added to public databases. It will be critical to conduct studies within various human populations if we are to fully understand how AIRR-seq can be leveraged to make improvements in a wide range of applications, including vaccine design.

Further, we suggest that the open AIRR-seq datasets could be leveraged to augment IG/TCR germline databases and inform the AIRR-seq and immunogenomics studies across diverse populations. It may be possible in the future to leverage AIRR-seq data to infer genetic ancestry, but such bioinformatics methods are yet to be developed. Conclusions about undocumented germline alleles based on non-targeted sequencing data including RNA-seq based on short read sequences should be considered with caution due to the complexity of the adaptive immune receptor loci[50,73], as described above. New methodologies and computational approaches may facilitate the inclusion of diverse population datasets into existing databases that should eventually reflect the genomic immunological diversity present in populations around the globe. Such enriched databases would provide researchers with the resources to address the next-generation design of personalized and precision immunodiagnostics and therapeutics[74].

Our interdisciplinary group consists of leading researchers from 17 regions including the USA, Canada, Norway, France, Sweden, the UK, Russia, Saudi Arabia, Israel, South Africa, Nigeria, Chile, Peru, China, Japan, Taiwan, and French Polynesia who share concerns about the lack of diversity in immunogenomics and embrace a need to tackle these challenges.

**Acknowledgments**

M.S. is supported by the Ministry of Science and Higher Education of the Russian Federation grant No. 075-15-2020-807. V.D.J. was supported by an award from the National Cancer Institute of the National Institutes of Health (K12CA001727). The laboratory of L.Q.-M. is supported by the Institut Pasteur, the Collège de France, the CNRS, the Fondation Allianz-Institut de France, and the French Government's Investissement d'Avenir program, Laboratoires d'Excellence 'Integrative Biology of Emerging Infectious Diseases' (ANR-10- LABX-62-IBEID) and 'Milieu Intérieur' (ANR-10-LABX-69-01). R.A.A. is supported by the Fogarty International Center of the National Institute of Health under award number D43TW010934. C.S. is supported by the National Institute of Allergy and Infectious Diseases of the National Institutes of Health under Award Number U01AI136677. G.K.H is supported by a grant from the Swedish Research Council (award number 532 2017-00968). V.G. is supported by a UiO World-Leading Research Community grant, the UiO:LifeScience Convergence Environment Immunolingo, EU Horizon 2020 iReceptorplus (#825821), and a Research Council of Norway FRIPRO project (#300740). We thank Dr. Nicky Mulder for the valuable comments that greatly improved the manuscript.

**Conflicts of interest statement**

V.G. declares advisory board positions in aiNET GmbH and Enpicom B.V. G.K.H and M.C. are founders of ImmuneDiscover Sweden AB.

# Supplementary table

Supplemental table 1. Tools for inference of germline TR/IG genes from AIRR-seq data

| Tool | Type of receptors | Type of inferring genes | Needs gene database for inference | Comment |
|---|---|---|---|---|
| TIgGER[38] | IG | V | yes | TIgGER and Partis assign AIRR-seq reads to V genes from the database and reports a list of V gene alleles (both known alleles and alleles with modifications). |
| Partis[39] | IG | V | yes | |
| IgDiscover[40] | IG / TCR | V, J | yes | IgDiscover uses the database for annotation of AIRR-seq reads, clusters reads with similar annotations, and reports both known and undocumented V genes. |
| IMPre[41] | IG / TCR | V, J | no | IMPre infers V and J genes from clusters of similar AIRR-seq reads and uses a germline database (if available) for annotation of the inferred genes. |
| IgScout[42] | IG | D, J | no | Both IgScout and MINING-D infer D genes as abundant substrings of CDR3s of AIRR-seq reads and use a germline database (if available) for annotation of the inferred genes. |
| MINING-D[43] | IG / TCR | D | no | |

References

1. Sirugo, G., Williams, S. M. & Tishkoff, S. A. The Missing Diversity in Human Genetic Studies. *Cell* **177**, 1080 (2019).

2. Ben-Eghan, C. *et al.* Don't ignore genetic data from minority populations. *Nature* **585**, 184–186 (2020).

3. McGuire, A. L. *et al.* The road ahead in genetics and genomics. *Nat. Rev. Genet.* (2020) doi:10.1038/s41576-020-0272-6.

4. Mills, M. C. & Rahal, C. A scientometric review of genome-wide association studies. *Commun Biol* **2**, 9 (2019).

5. Sherman, R. M. *et al.* Assembly of a pan-genome from deep sequencing of 910 humans of African descent. *Nat. Genet.* **51**, 30–35 (2019).

6. Gurdasani, D. *et al.* The African Genome Variation Project shapes medical genetics in Africa. *Nature* **517**, 327–332 (2015).

7. Choudhury, A. *et al.* Whole-genome sequencing for an enhanced understanding of genetic variation among South Africans. *Nat. Commun.* **8**, 2062 (2017).

8. Choudhury, A. *et al.* High-depth African genomes inform human migration and health. *Nature* **586**, 741–748 (2020).

9. Retshabile, G. *et al.* Whole-Exome Sequencing Reveals Uncaptured Variation and Distinct Ancestry in the Southern African Population of Botswana. *Am. J. Hum. Genet.* **102**, 731–743 (2018).

10. GenomeAsia100K Consortium. The GenomeAsia 100K Project enables genetic discoveries across Asia. *Nature* **576**, 106–111 (2019).

11. Mulder, N. *et al.* H3Africa: current perspectives. *Pharmgenomics. Pers. Med.* **11**, 59–66 (2018).

12. H3Africa Consortium *et al.* Research capacity. Enabling the genomic revolution in Africa. *Science* **344**, 1346–1348 (2014).

13. Lin, J.-C., Hsiao, W. W.-W. & Fan, C.-T. Transformation of the Taiwan Biobank 3.0: vertical and

horizontal integration. *J. Transl. Med.* **18**, 304 (2020).

14. Wojcik, G. L. *et al.* Genetic analyses of diverse populations improves discovery for complex traits. *Nature* **570**, 514–518 (2019).

15. Amendola, L. M. *et al.* The Clinical Sequencing Evidence-Generating Research Consortium: Integrating Genomic Sequencing in Diverse and Medically Underserved Populations. *Am. J. Hum. Genet.* **103**, 319–327 (2018).

16. All of Us Research Program Investigators *et al.* The 'All of Us' Research Program. *N. Engl. J. Med.* **381**, 668–676 (2019).

17. Manser, A. R., Weinhold, S. & Uhrberg, M. Human KIR repertoires: shaped by genetic diversity and evolution. *Immunol. Rev.* **267**, 178–196 (2015).

18. Gourraud, P.-A. *et al.* HLA diversity in the 1000 genomes dataset. *PLoS One* **9**, e97282 (2014).

19. Freeman, J. D., Warren, R. L., Webb, J. R., Nelson, B. H. & Holt, R. A. Profiling the T-cell receptor beta-chain repertoire by massively parallel sequencing. *Genome Res.* **19**, 1817–1824 (2009).

20. Weinstein, J. A., Jiang, N., White, R. A., 3rd, Fisher, D. S. & Quake, S. R. High-throughput sequencing of the zebrafish antibody repertoire. *Science* **324**, 807–810 (2009).

21. Boyd, S. D. *et al.* Individual variation in the germline Ig gene repertoire inferred from variable region gene rearrangements. *J. Immunol.* **184**, 6986–6992 (2010).

22. Tumeh, P. C. *et al.* PD-1 blockade induces responses by inhibiting adaptive immune resistance. *Nature* **515**, 568–571 (2014).

23. Han, A., Glanville, J., Hansmann, L. & Davis, M. M. Linking T-cell receptor sequence to functional phenotype at the single-cell level. *Nat. Biotechnol.* **32**, 684–692 (2014).

24. Briney, B., Inderbitzin, A., Joyce, C. & Burton, D. R. Commonality despite exceptional diversity in the baseline human antibody repertoire. *Nature* **566**, 393–397 (2019).

25. Brown, A. J. *et al.* Augmenting adaptive immunity: progress and challenges in the quantitative engineering and analysis of adaptive immune receptor repertoires. *Molecular Systems Design & Engineering* vol. 4 701–736 (2019).

26. Avnir, Y. *et al.* IGHV1-69 polymorphism modulates anti-influenza antibody repertoires, correlates with IGHV utilization shifts and varies by ethnicity. *Sci. Rep.* **6**, 20842 (2016).

27. Setliff, I. *et al.* Multi-Donor Longitudinal Antibody Repertoire Sequencing Reveals the Existence of Public Antibody Clonotypes in HIV-1 Infection. *Cell Host Microbe* **23**, 845–854.e6 (2018).

28. Roskin, K. M. *et al.* Aberrant B cell repertoire selection associated with HIV neutralizing antibody breadth. *Nat. Immunol.* **21**, 199–209 (2020).

29. Kreer, C., Gruell, H., Mora, T., Walczak, A. M. & Klein, F. Exploiting B Cell Receptor Analyses to Inform on HIV-1 Vaccination Strategies. *Vaccines (Basel)* **8**, (2020).

30. Emerson, R. O. *et al.* Immunosequencing identifies signatures of cytomegalovirus exposure history and HLA-mediated effects on the T cell repertoire. *Nature Genetics* vol. 49 659–665 (2017).

31. Liu, X. S. & Mardis, E. R. Applications of Immunogenomics to Cancer. *Cell* **168**, 600–612 (2017).

32. Sharonov, G. V., Serebrovskaya, E. O., Yuzhakova, D. V., Britanova, O. V. & Chudakov, D. M. B cells, plasma cells and antibody repertoires in the tumour microenvironment. *Nat. Rev. Immunol.* **20**, 294–307 (2020).

33. Simnica, D. *et al.* High-Throughput Immunogenetics Reveals a Lack of Physiological T Cell Clusters in Patients With Autoimmune Cytopenias. *Front. Immunol.* **10**, 1897 (2019).

34. Bashford-Rogers, R. J. M. *et al.* Analysis of the B cell receptor repertoire in six immune-mediated diseases. *Nature* vol. 574 122–126 (2019).

35. Liu, X. *et al.* T cell receptor β repertoires as novel diagnostic markers for systemic lupus erythematosus and rheumatoid arthritis. *Ann. Rheum. Dis.* **78**, 1070–1078 (2019).

36. Fraussen, J. *et al.* Phenotypic and Ig Repertoire Analyses Indicate a Common Origin of IgDCD27 Double Negative B Cells in Healthy Individuals and Multiple Sclerosis Patients. *J. Immunol.* **203**, 1650–1664 (2019).

37. Gate, D. *et al.* Clonally expanded CD8 T cells patrol the cerebrospinal fluid in Alzheimer's disease. *Nature* vol. 577 399–404 (2020).

38. Gadala-Maria, D. *et al.* Identification of Subject-Specific Immunoglobulin Alleles From Expressed

Repertoire Sequencing Data. *Front. Immunol.* **10**, 129 (2019).

39. Ralph, D. K. & Matsen, F. A., IV. Per-sample immunoglobulin germline inference from B cell receptor deep sequencing data. *PLoS Comput. Biol.* **15**, e1007133 (2019).

40. Corcoran, M. M. *et al.* Production of individualized V gene databases reveals high levels of immunoglobulin genetic diversity. *Nat. Commun.* **7**, 13642 (2016).

41. Zhang, W. *et al.* IMPre: An Accurate and Efficient Software for Prediction of T- and B-Cell Receptor Germline Genes and Alleles from Rearranged Repertoire Data. *Frontiers in Immunology* vol. 7 (2016).

42. Safonova, Y. & Pevzner, P. A. De novo Inference of Diversity Genes and Analysis of Non-canonical V(DD)J Recombination in Immunoglobulins. *Frontiers in Immunology* vol. 10 (2019).

43. Bhardwaj, V., Franceschetti, M., Rao, R., Pevzner, P. A. & Safonova, Y. Automated analysis of immunosequencing datasets reveals novel immunoglobulin D genes across diverse species. *PLoS Comput. Biol.* **16**, e1007837 (2020).

44. Ohlin, M. *et al.* Inferred Allelic Variants of Immunoglobulin Receptor Genes: A System for Their Evaluation, Documentation, and Naming. *Front. Immunol.* **10**, 435 (2019).

45. Mikocziova, I. *et al.* Polymorphisms in human immunoglobulin heavy chain variable genes and their upstream regions. *Nucleic Acids Res.* **48**, 5499–5510 (2020).

46. Scheepers, C. *et al.* Ability To Develop Broadly Neutralizing HIV-1 Antibodies Is Not Restricted by the Germline Ig Gene Repertoire. *The Journal of Immunology* vol. 194 4371–4378 (2015).

47. Wang, Y. *et al.* Genomic screening by 454 pyrosequencing identifies a new human IGHV gene and sixteen other new IGHV allelic variants. *Immunogenetics* **63**, 259–265 (2011).

48. Watson, C. T. *et al.* Complete haplotype sequence of the human immunoglobulin heavy-chain variable, diversity, and joining genes and characterization of allelic and copy-number variation. *Am. J. Hum. Genet.* **92**, 530–546 (2013).

49. Zhang, J.-Y. *et al.* Using de novo assembly to identify structural variation of complex immune system gene regions. *Cold Spring Harbor Laboratory* 2021.02.03.429586 (2021)

doi:10.1101/2021.02.03.429586.

50. Rodriguez, O. L. *et al.* A Novel Framework for Characterizing Genomic Haplotype Diversity in the Human Immunoglobulin Heavy Chain Locus. *Frontiers in Immunology* **11**, 2136 (2020).

51. Lefranc, M.-P. *et al.* IMGT®, the international ImMunoGeneTics information system® 25 years on. *Nucleic Acids Res.* **43**, D413–22 (2015).

52. Christley, S. *et al.* The ADC API: A Web API for the Programmatic Query of the AIRR Data Commons. *Front. Big Data* **3**, (2020).

53. Omer, A. *et al.* VDJbase: an adaptive immune receptor genotype and haplotype database. *Nucleic Acids Res.* **48**, D1051–D1056 (2020).

54. Brzezinski, J. L., Deka, R., Menon, A. G., Glass, D. N. & Choi, E. Variability in TRBV haplotype frequency and composition in Caucasian, African American, Western African and Chinese populations. *Int. J. Immunogenet.* **32**, 413–420 (2005).

55. Mackelprang, R. *et al.* Sequence diversity, natural selection and linkage disequilibrium in the human T cell receptor alpha/delta locus. *Human Genetics* vol. 119 255–266 (2006).

56. Mackelprang, R. *et al.* Sequence variation in the human T-cell receptor loci. *Immunological Reviews* vol. 190 26–39 (2002).

57. Calonga-Solís, V. *et al.* Unveiling the Diversity of Immunoglobulin Heavy Constant Gamma () Gene Segments in Brazilian Populations Reveals 28 Novel Alleles and Evidence of Gene Conversion and Natural Selection. *Front. Immunol.* **10**, 1161 (2019).

58. Romo-González, T., Morales-Montor, J., Rodríguez-Dorantes, M. & Vargas-Madrazo, E. Novel substitution polymorphisms of human immunoglobulin VH genes in Mexicans. *Hum. Immunol.* **66**, 732–740 (2005).

59. Watson, C. T. & Breden, F. The immunoglobulin heavy chain locus: genetic variation, missing data, and implications for human disease. *Genes Immun.* **13**, 363–373 (2012).

60. Watson, C. T., Glanville, J. & Marasco, W. A. The Individual and Population Genetics of Antibody Immunity. *Trends Immunol.* **38**, 459–470 (2017).

61. Quintana-Murci, L. Human Immunology through the Lens of Evolutionary Genetics. *Cell* **177**, 184–199 (2019).

62. Kruszka, P. *et al.* Williams-Beuren syndrome in diverse populations. *Am. J. Med. Genet. A* **176**, 1128–1136 (2018).

63. Jaratlerdsiri, W. *et al.* Whole-Genome Sequencing Reveals Elevated Tumor Mutational Burden and Initiating Driver Mutations in African Men with Treatment-Naïve, High-Risk Prostate Cancer. *Cancer Res.* **78**, 6736–6746 (2018).

64. Ioannidis, J. P. A., Powe, N. R. & Yancy, C. Recalibrating the Use of Race in Medical Research. *JAMA* (2021) doi:10.1001/jama.2021.0003.

65. Bauchner, H. & Fontanarosa, P. B. Randomized Clinical Trials and COVID-19: Managing Expectations. *JAMA* **323**, 2262–2263 (2020).

66. Initiative, T. C.-19 H. G. & The COVID-19 Host Genetics Initiative. The COVID-19 Host Genetics Initiative, a global initiative to elucidate the role of host genetic factors in susceptibility and severity of the SARS-CoV-2 virus pandemic. *European Journal of Human Genetics* vol. 28 715–718 (2020).

67. Ellinghaus, D. *et al.* Genomewide Association Study of Severe Covid-19 with Respiratory Failure. *N. Engl. J. Med.* (2020) doi:10.1056/NEJMoa2020283.

68. Huang, J. *et al.* Identification of a CD4-Binding-Site Antibody to HIV that Evolved Near-Pan Neutralization Breadth. *Immunity* **45**, 1108–1121 (2016).

69. Robbiani, D. F. *et al.* Recurrent Potent Human Neutralizing Antibodies to Zika Virus in Brazil and Mexico. *Cell* **169**, 597–609.e11 (2017).

70. Yuan, M. *et al.* Structural basis of a shared antibody response to SARS-CoV-2. *Science* **369**, 1119–1123 (2020).

71. Havenar-Daughton, C. *et al.* The human naive B cell repertoire contains distinct subclasses for a germline-targeting HIV-1 vaccine immunogen. *Sci. Transl. Med.* **10**, (2018).

72. Burton, D. R. Advancing an HIV vaccine; advancing vaccinology. *Nat. Rev. Immunol.* **19**, 77–78 (2019).

73. Watson, C. T. *et al.* Comment on 'A Database of Human Immune Receptor Alleles Recovered from Population Sequencing Data'. *The Journal of Immunology* vol. 198 3371–3373 (2017).

74. Mining adaptive immune receptor repertoires for biological and clinical information using machine learning. *Current Opinion in Systems Biology* **24**, 109–119 (2020).